\documentclass{article}

\usepackage{graphicx}

\usepackage[final]{neurips_2024}

\usepackage[utf8]{inputenc} 
\usepackage[T1]{fontenc}    
\usepackage{hyperref}       
\usepackage{url}            
\usepackage{booktabs}       
\usepackage{amsfonts}       
\usepackage{nicefrac}       
\usepackage{microtype}      
\usepackage{xcolor}         
\usepackage{amsmath}
\usepackage{paralist}
\usepackage{comment}
\usepackage[affil-it]{authblk}
\title{DreamLLM-3D: Affective Dream Reliving using Large Language Model and 3D Generative AI}

\author[1]{Pinyao Liu}
\author[2]{Keon Ju Lee}
\author[1]{Alexander Steinmaurer}
\author[3]{Claudia Picard-Deland}
\author[3]{Michelle Carr}
\author[2]{Alexandra Kitson}

\affil[1]{Interdisciplinary Transformation University Austria}
\affil[2]{Simon Fraser University}
\affil[3]{Université de Montréal}

\begin{document}

\maketitle

\begin{figure}[h]
\centering
\includegraphics[width=\textwidth]{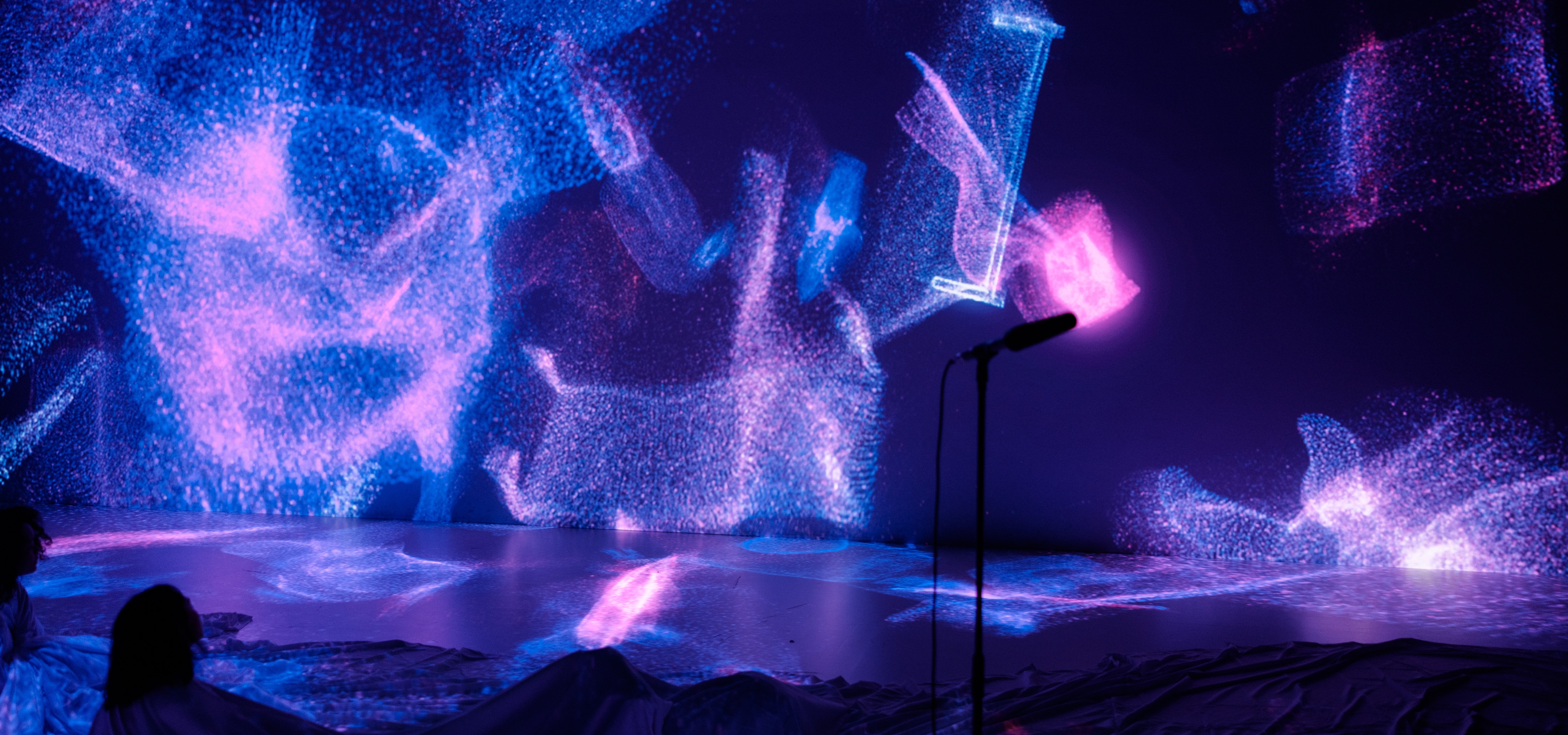}
\caption{During the dream-reliving experience, the immersant whispers their dream into the system. The spoken dream characters and objects are then represented as 3D point clouds in the shared immersive space.}
\label{fig:teaser}
\end{figure}
\begin{abstract}
We present DreamLLM-3D, a composite multimodal AI system behind an immersive art installation for dream re-experiencing. It enables automated dream content analysis for immersive dream-reliving, by integrating a Large Language Model (LLM) with text-to-3D Generative AI. The LLM processes voiced dream reports to identify key dream entities (characters and objects), social interaction, and dream sentiment. The extracted entities are visualized as dynamic 3D point clouds, with emotional data influencing the color and soundscapes of the virtual dream environment. Additionally, we propose an experiential AI-Dreamworker Hybrid paradigm. Our system and paradigm could potentially facilitate a more emotionally engaging dream-reliving experience, enhancing personal insights and creativity. 
\end{abstract}

\section{Introduction}

Dreams have a long history of inspiring creativity. Salvador Dalí's iconic melting clocks were inspired by a dream that embodies the fluidity of time and reality, while Niels Bohr discovered the structure of the atom and credited a dream where he saw the electrons revolving around the nucleus like the solar system.  Attaining personal insights from interpreting dreams has been the subject of extensive interest for centuries. Psychologists and neuroscientists have put forth several "dreamwork" strategies that support people in deeply engaging with their dreams \citep{hill_use_2010}. One of the more salient strategies includes re-experiencing the dream as if reliving the memory, feelings, and bodily sensations from the dream \citep{ellis_common_2019}.

In addition to the experiential approach of dreamwork, a more quantitative approach has been used for dream content analysis. One of the most widely used dream content analysis systems is the Hall–Van de Castle (HVDC) scheme \citep{hall_content_1966}. It consists of ten categories of elements appearing in dreams (see Section 2), together with detailed rules to recognize and measure those elements from written reports. 


In dream content analysis, manually scoring dreams according to coding rules can be very time-consuming \citep{fogli_our_2020}. AI, therefore, could be helpful to automate the process. More and more dream researchers have been deploying automation and AI for dream analysis -- such as automatically scoring dream reports by operationalizing the HVDC coding rules \citep{fogli_our_2020}, mapping dream contents for semantic structures of characters, activities, and setting using a transformer based Natural Language Processing architecture \citep{gutman_music_mapping_2022}, and automatic scoring of dream reports with LLMs \citep{bertolini_automatic_2023,picard-deland_c_vestibular_2023}. These AI-assisted dream content analyses have mainly been used by researchers to discover themes and patterns among a vast database of dream reports but are rarely used in the dreamwork context to attain personal insights.

Relevant to experiential dreamwork, researchers are exploring AI for dream imagery visualization and re-experiencing. Some directly look at how AI can help decode fMRI signals and reconstruct visual images \citep{miyawaki_visual_2008,kay_identifying_2008,takagi_high-resolution_2023}, and further decode visual imagery during sleep \citep{horikawa_neural_2013} and dreamed objects \citep{horikawa_hierarchical_2017}. Other researchers have started to explore Generative AI for dream image visualization, mostly using text-to-2D AI models to visualize dreamers' dream record \citep{canet_sola_dream_2022,fuse_onirica_nodate,liu_falling_2024}. More recently, \citet{liu_virtual_2024} explored text-to-3D AI for re-experiencing dreams through transforming whispered dream speech into 3D dream objects in Virtual Reality. However, this dream-reliving system does not account for the rich emotional and social information from the dream narration. Emotions are suggested an important component of most dream experiences and may enhance dream recall \citep{blagrove_relationship_2004}. Moreover, Social Simulation Theory (SST) postulates that dreams virtually simulate socially significant interactions, bonds, and networks during our waking lives \citep{revonsuo_avatars_2015}. By incorporating emotional and social information into the immersive re-experiencing of dream imagery, we could potentially allow for a more affective dream-reliving experience for deepened personal insights.

In summary, we see the potential of AI-assisted dream content analysis to enhance experiential dreamwork. We propose to bridge the divide between the experiential and the analytical with a multimodal AI system that combines dream content analysis and dream imagery re-experiencing, using text-to-3D generative diffusion model and LLM. We further propose an experiential AI-Dreamworker Hybrid paradigm and discuss its ethical implications. Our system and paradigm could potentially facilitate a more emotionally engaging dream-reliving experience, supporting deeper personal insights, while provoking meaningful discussions within dream and AI research communities. 

\section{AI-assisted Affective Dream Reliving Overview}
\begin{figure}
    \centering
    \includegraphics[width=\textwidth]{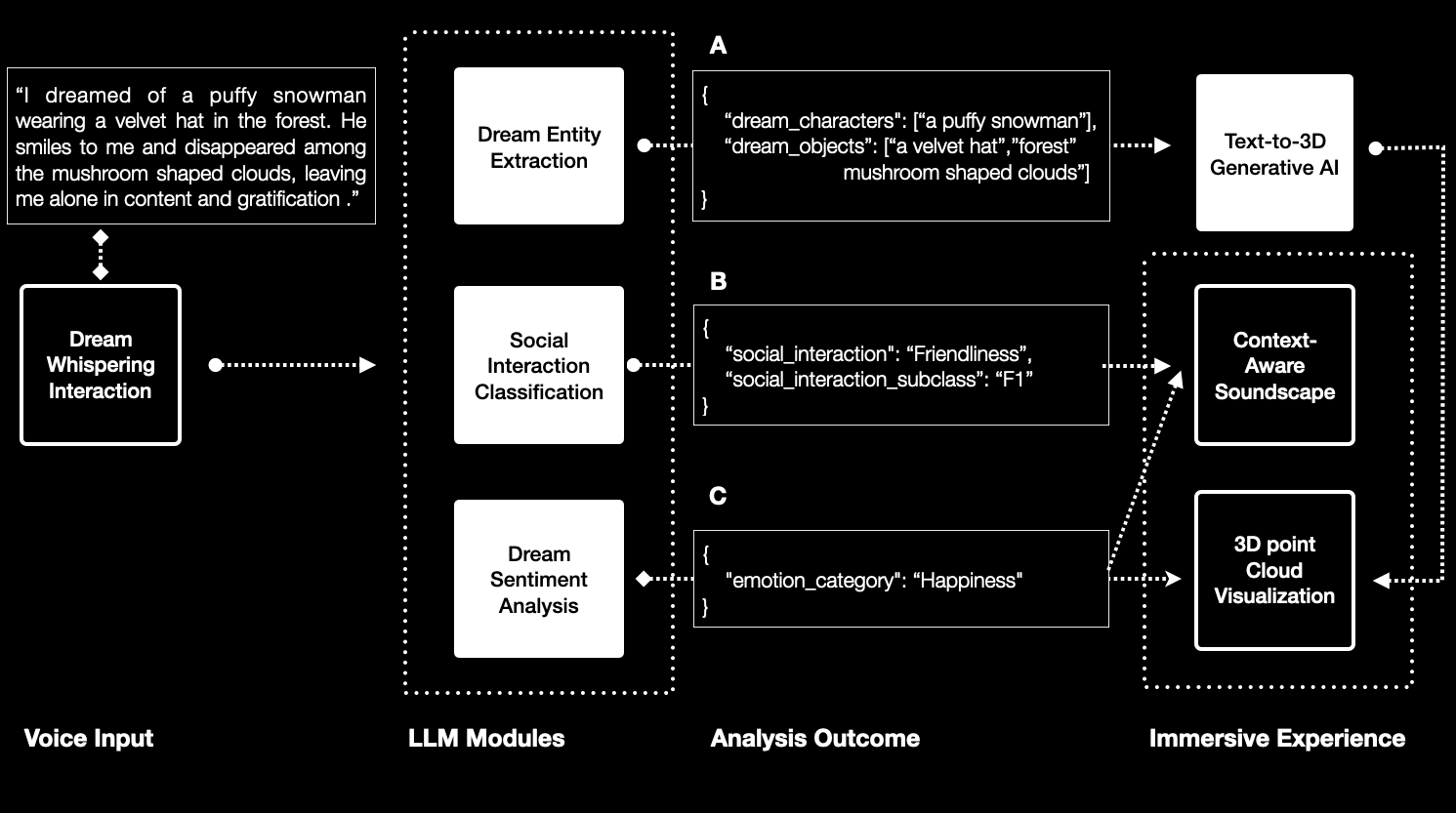}
    \caption{A composite multimodal AI system performing real-time dream content analysis for immersive dream-reliving, using LLM and 3D generative AI. }
    \label{fig:mechanism}
\end{figure}
We developed an LLM-assisted dream content analysis system behind an immersive art installation \textbf{for} an affective dream-reliving experience inspired by the HVDC scheme. The HVDC scheme consists of ten categories: 
\begin{inparaenum}
  \item Characters
  \item Social Interactions
  \item Emotions
  \item Activities (walking, talking, thinking, etc.)
  \item Success and Failure
  \item Misfortune and Good Fortune
  \item Settings and Objects
  \item Descriptive Elements (modifiers, time, negatives)
  \item Food and Eating
  \item Elements from the Past
\end{inparaenum} 
\citep{hall_content_1966}. \citet{domhoff_scientific_2003} suggests that the three categories of characters, social interactions and emotions are the most valuable and informative, which we thus chose to output as structured information from each dream.

During the dream-reliving experience, the immersant whispers their dream into the system through a microphone (see Figure \ref{fig:teaser}). In each whispering session, the system activates speech recognition when detecting voice input and stops when sensing a pause or stop in the vocal narration. Each whispering input usually contains one or two sentences of a dream report. An LLM pipeline analyzes and categorizes key dream entities, sentiments, and social interaction in real-time. The key dream entity information is output as automatized prompts for the text-to-3D diffusion model. The social interaction information is mapped into the soundscape, while the sentiment information is mapped into both the visual (color and motion of 3D generative visualization) and the soundscape. The immersant then re-experiences one's dream memories through AI-generated 3D dream objects and characters, with affective color mapping and context-aware soundscape.

Below we detail the three LLM dream analysis modules, and their integration with the text-to-3D model. We further detail the affective color mapping, and the context-aware soundscape system. 
\subsection{LLM System}

\textbf{Technical Setup}
We used a local LLM Mistral 7B \citep{jiang_mistral_2023} with temperature $T = 0$. Additionally, we used Nomic-Embed-Text \citep{nussbaum_nomic_2024} as our embedding model. 
We then used Chroma as the search library, which natively supports embedding normalization and cosine similarity search, since we are interested in the association of terms with social interactions (the direction of vectors).

We used a generative diffusion text-to-3D model \textit{Point-E} \citep{nichol_point-e_2022} for dream imagery visualization, which produces a 3D point cloud representation of dream entities (characters and objects) in about 17 seconds on an A100 GPU. These 3D point clouds are then parsed in the Unity3D engine for real-time rendering.  


\textbf{Dream Entity Extraction Module}
The HVDC scheme only classifies living entities as characters, and views non-living dream objects as tangible 'props' - part of the physical environment that these living entities interact with \citep{hall_content_1966}. However, Hall also pointed out that "A dream symbol is an image, usually a visual image, of an object, activity, or scene; the referent for the symbol is a conception" \citep{hall_cognitive_1953}. This suggests the potential significant symbolic meaning attached to a non-living entity in dreams. Therefore, simplifying dream objects as physical props might lead to the loss of significant symbolic meanings in an experiential dreamwork approach. Thus, we placed the significance of non-living objects as equal to that of living humans/animals/fictional figures and used the term \textit{"dream entity"} to encompass both living dream characters and non-living dream objects. 

One challenge with the \textit{point-E} model is that, in the pure text-conditional generation, it grapples with prompts that combine multiple entities and works better with simple prompts that describe a single entity. Thus, we designed the dream entity extraction module to perform segmentation for each dream whispering interaction. It breaks down a complex dream scene description into automized prompts of single dream entities, tailored to serve the text-to-3D generative diffusion model (see Figure \ref{fig:mechanism} A). 


\textbf{Social Interaction Classification Module}
The second component of the LLM analyzes the social interactions among the dream entities. According to the HVDC scheme, social interaction in dreams was coded into three classes: aggressive, friendly, and sexual interactions. We used the definitions of each class and its subclass as embeddings, which categorize Aggressions into 8 levels from A1 (Covert hostility or anger) to A8 (Death by aggression), Friendliness into 7 levels from F1 (Felt but unexpressed friendliness) to F7 (Long-term relationship), and Sexual Interactions into 5 levels from S1 (Sexual thoughts or fantasies) to S5 (Sexual intercourse or attempt). 
The LLM module outputs one dominant social interaction subclass for each whispering interaction (see Figure \ref{fig:mechanism} B).





\textbf{Dream Sentiment Analysis Module}
The third component of the LLM classifies representative dream emotions. According to the HVDC scheme, dream emotions were coded into five classes: Anger (AN), Apprehension (AP), Sadness (SD), Confusion (CO), and Happiness (HA). The LLM module analyzes the dream speech and outputs one dominant emotion class for each whispering interaction (see Figure \ref{fig:mechanism} C).

\subsection{Affective Color Mapping}
The dream sentiment output further influences the color mapping of the AI-generated 3D point clouds during real-time rendering. We refer to affective color psychology and map each dream emotion category to a single representative color, based on cross-cultural studies on color and emotion \citep{madden_managing_2000,bartram_affective_2017}. We also situate the emotions within Russell's Circumplex model of affect \citep{russell_circumplex_1980} and map the dream emotions into single representative colors and their relative positions on the Valence and Arousal coordinate.
We created subclasses for Happiness to map peacefulness (high Valence, low Arousal) into light blue, and excitement (High Valence, high Arousal) into gold. We mapped Anger (low Valence, high Arousal) into dark red, and Apprehension (low Valence, neutral Arousal) into purple, Sadness (low Valence, low Arousal) into dark blue. Not seen as a typical emotion, confusion (neutral Valence, neutral to high Arousal) was less mentioned in affective color theory. The HVDC scheme refers to confusion as a fleeting state of "cognitive ambiguity" when confronting the unexpected or uncertain \citep{hall_content_1966}. We chose to map confusion into white glow in the artwork using the metaphor of fog.

\begin{figure}
  \centering
  \includegraphics[width=\textwidth]{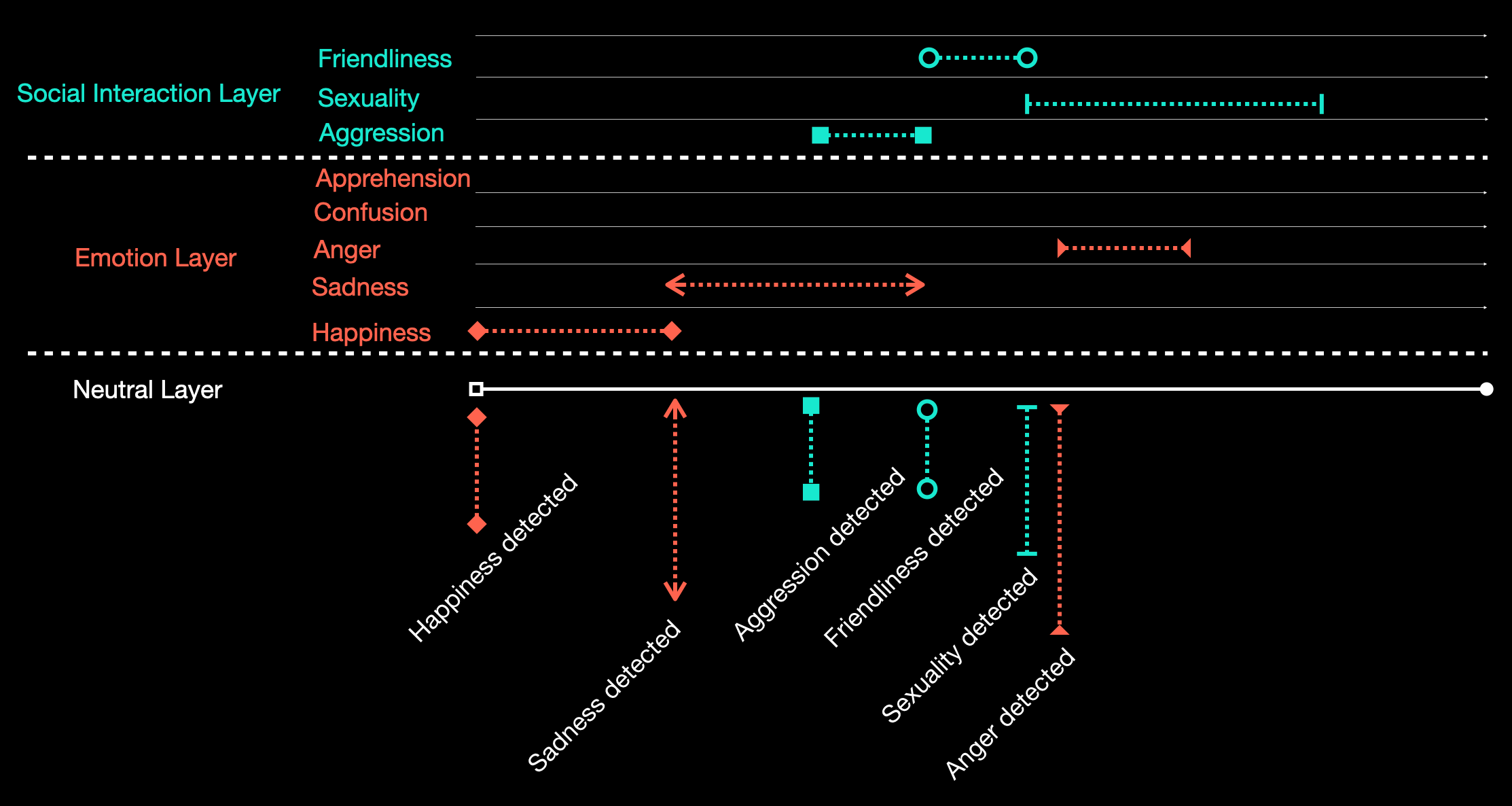}
  \caption{Example of a graphic score for soundscape composition illustrating the integration of sound layers for the emotion classes and social interaction classes (X-axis = time, and Y-axis = sound layers).}
  \label{fig:graphical-note}
\end{figure}
\subsection{Context-Aware Soundscape System}
We designed a context-aware soundscape system, interacting with both social interaction and dream sentiment output, to enhance the affective experience. First, the music composer composed a neutral sound layer as the base layer for a composite soundscape. Then, he composed a single sound layer for each specific emotional class and social interaction class. During the experience, the system detected the emotion class and social interaction class in real-time and adjusted the composite sound layers accordingly. For instance, as depicted in Figure \ref{fig:graphical-note} with a graphic score, when the emotion class \textit{happiness}  is detected, the happiness sound layer is integrated with the neutral layer. And then, when the emotion class \textit{sadness} is detected, the sadness sound layer substitutes the happiness sound layer. Additionally, if a social interaction class \textit{aggression} is identified, an aggression sound layer blends on top of the emotion sound layer.

\section{Discussion and Conclusion}
Our contributions are twofold: first, we implemented a composite AI System that extracts affective dream information for immersive dream-reliving experiences, using the Large Language Model and text-to-3D Generative AI model; second, we propose an experiential AI-dreamworker hybrid paradigm as a future interface for dreamwork.

\subsection{A Composite Multimodal AI System for Affective Dream Re-experiencing.} 




Through our immersive art installation, we implemented an automated dream sentiment and social interaction analysis enabled by LLM, with the output data influencing a multisensory immersive dream-reliving experience. Within AI-assisted dream content analysis, \citet{bertolini_automatic_2023} are the first to use language models to predict the absence or presence of emotions, while \citet{cortal_sequence--sequence_2024} use a sequence-to-sequence language model to automate the annotation of both characters and their emotions. These approaches both follow the HVDC scheme, but focus only on pattern finding within a large database rather than individual dreamwork. While in the experiential dreamwork approach, previous work focuses on text-to-image generation based on the dream report, without considering the rich affective information from dream narration (dream sentiment and social interaction) \citep{fuse_onirica_nodate,canet_sola_dream_2022,liu_falling_2024}.

To the best of our knowledge, our work represents the first attempt to integrate on-the-fly content analysis with experiential dreamwork, bridging the long-existing divide between the experiential \citep{hill_use_2010,ellis_diving_2023} and the analytical (e.g. HVDC). The continuity hypothesis posits that most dreams are a continuation of what is happening in everyday life and that patterns of dream content reflect the dreamer’s most important waking life concerns, interests, and activities \citep{hall_content_1966,domhoff_scientific_2003}. Thus, by integrating AI-assisted dream content extraction into experiential dreamwork, we could potentially provide the dreamer with more personal insights, and dream researchers with a new direction in integrating the analytical data into the nuanced phenomenological experience of the dreamer.
In the future, we plan to adapt the experience into a longitudinal re-experiencing and analysis tool and further evaluate its potential in facilitating affective dream re-experiencing and insight gain.

\subsection{Towards an Experiential AI-Dreamworker Hybrid Paradigm} 
The role of AI in the field of dream research has evolved through distinct phases, each contributing unique capabilities to the analysis and interpretation of dreams:

\textbf{AI as a Textual Dream Interpreter:} These systems analyze dream reports through either automating content analysis based on specific coding rules \citep{gutman_music_mapping_2022,cortal_sequence--sequence_2024,bertolini_automatic_2023},  thematic analyses \citep{picard-deland_c_vestibular_2023}, or association-based interpretation \citep{kelly_bulkeley_freud_2024}. While these approaches effectively identify key themes, entities, and narratives, their primary focus has been on static analysis rather than engaging with the experiential aspects of dreaming or dreamwork.

\textbf{AI as a Visuospatial Dream Interpreter:}
Most previous artworks and studies focus on text-to-image for dream visualization \citep{canet_sola_dream_2022,fuse_onirica_nodate,liu_falling_2024}. The recent incorporation of 3D generative models has transformed AI into a visuospatial interpreter, allowing users to explore dream imagery spatially for deepened experiential engagement \citep{liu_virtual_2024}. 

\textbf{Experiential AI-Dreamworker Hybrid:}
We propose the concept of an AI-dreamworker hybrid, which will represent a significant advancement in this trajectory. The AI dream worker will not only extract and analyze affective information from dreams—such as emotions and social interactions as per our current system—but will also take this a step further by \textbf{observing} the dreamer’s behavior during the dream reliving process, such as in a Virtual Reality environment with a Head Mounted Display. The soundscape and dream imagery could be transformed based on the interaction between the dreamer and the dream entities and environment (e.g. proximity and hand interaction). Moreover, the hybrid AI dreamworker system can adaptively \textbf{guide} the dreamer's attention by highlighting a specific dream entity or even \textbf{suggesting} ways to interact with the dream objects with a voice, leveraging established experiential methods (e.g. embodying a dream element, dreaming the dream onward, prompting dialogue with a dream character, or focusing on specific emotions or bodily sensations \citep{ellis_common_2019,ellis_diving_2023}). By integrating affective knowledge and real-time observation, the AI will become a more interactive and responsive participant in the dreamwork process, potentially aiding in deeper self-reflection and insight. Meanwhile, we envision that the human dreamworker, equipped with intuition and a deeper understanding of the subjective nuances of dreaming, will collaborate with AI to guide and enrich the interpretive process. The AI's guidance and extraction of affective information serve as a foundation, while the human dreamworker can offer personalized facilitation and interpretation, creating a holistic approach to dreamwork that balances computational precision with human empathy and creativity.

\subsection{Ethical implication:} 
While AI makes the dreamwork and entry-level dream analysis more accessible and affordable, it also introduces ethical challenges that require careful consideration. One risk is that the overuse of AI dreamwork and analysis systems might lead to \textbf{isolation and disconnection} to human communities of dream-sharing, which is crucial to meaning-making and community-finding for the dreamer. Therefore, it's essential to integrate AI in ways that preserve and enhance human-to-human interactions in dream-sharing. In addition, AI systems are susceptible to \textbf{biases} related to gender, race, and cultural stereotypes, which can lead to misleading interpretations or reinforce harmful narratives. These biases stem from the large datasets on which both language models and text-to-3D generative models are trained. Addressing these risks requires ongoing efforts in curating datasets, refining models, and critically evaluating AI outputs. Finally, the dreamer or dreamworker may become overly reliant on AI-generated guidance or interpretation, and \textbf{lose one's agency} in determining a dream's meaning. It is crucial to ensure that AI remains a tool to support, rather than dictate, the interpretive process, keeping the dreamer and dreamworker as active participants.

In \textit{The Apology of Socrates} \citep{plato_apology_1901}, Socrates describes his \textit{"Daimonion"} as an inner voice or divine sign:
\textit{"This sign, which is a kind of voice, first began to come to me when I was a child; it often holds me back (from doing something wrong or harmful to himself) but never commands me to do anything."} In essence, we envision a modern equivalent of Socrates' \textit{"Daimonion"} in dreamwork through AI — a voice that gently guides dream exploration, supporting the dreamer's journey without dictating their path, while fostering human insights and creativity flourishment.

\bibliography{NeurIPS2024Bib}




\end{document}